\documentclass[12pt,preprint]{aastex}
\def\bmath#1{\mbox{\boldmath$#1$}}
\begin{document}

\title{
Markov Chain Monte Carlo joint analysis of {\it Chandra} X-ray 
imaging spectroscopy and
Sunyaev-Zeldovich Effect data}


\author{Massimiliano~Bonamente$\,^{1,2}$, 
Marshall~K.~Joy$\,^{2}$, John~E.~Carlstrom$\,^{3,4}$, Erik~D.~Reese$\,^{5,6}$
and Samuel~J.~LaRoque$\,^{3,4}$
}

\affil{\(^{\scriptstyle 1} \){Department of Physics, University of Alabama,
Huntsville, AL}\\
\(^{\scriptstyle 2} \){NASA Marshall Space Flight Center, Huntsville, AL}\\
\(^{\scriptstyle 3} \){Department of Astronomy and Astrophysics, University of Chicago, Chicago, IL 60637}\\
\(^{\scriptstyle 4} \){Kavli Institute for Cosmological Physics, Department of Physics, 
Enrico Fermi Institute, University of Chicago, Chicago, IL 60637}\\
\(^{\scriptstyle 5} \){Physics Department, University of California, Berkeley, CA 94720}\\
\(^{\scriptstyle 6} \){Chandra Fellow}
}
\begin{abstract}
X-ray and Sunyaev-Zeldovich Effect data can be combined
to determine the distance to galaxy clusters.
High-resolution X-ray data are now available from the {\it Chandra} Observatory, which
provides both spatial and spectral information, and
Sunyaev-Zeldovich Effect data were obtained from the
BIMA and OVRO arrays.
We introduce a Markov chain Monte Carlo  procedure for the joint analysis 
of X-ray and Sunyaev-Zeldovich Effect data.
The advantages of this method  are the high computational efficiency and the ability
to measure simultaneously the probability distribution of all parameters of interest, such as the
spatial and spectral properties of the cluster gas and also
for derivative quantities such as the distance to the cluster.
We demonstrate this technique by applying it to the {\it Chandra} X-ray data and the OVRO radio data
for the galaxy cluster Abell~611. Comparisons with traditional likelihood-ratio methods
reveal the robustness of the method.
This method will be used in  follow-up papers to determine the distances to a large sample of
galaxy clusters.
\end{abstract}


\section{Introduction}
Analysis of  Sunyaev-Zeldovich Effect (SZE) and X-ray data provides
a unique method of directly determining distances of galaxy clusters.
Clusters of galaxies contain hot plasma ($k_B T_e \sim$ 2-20 keV) that
scatters the cosmic microwave background radiation (CMB). On average,  this inverse Compton 
scattering boosts the energy of the CMB photons, causing a small distortion in
the CMB spectrum (Sunyaev and Zeldovich 1970,1972).
For recent reviews of the SZE and its application for
cosmology see Birkinshaw (1999) and Carlstrom et al. (2002).

The SZE is proportional to the integrated pressure along the line of sight,
$\Delta T \propto  \int n_e T_e dl$, where $n_e$ and $T_e$ are
the electron density and electron temperature of the cluster plasma.
The thermal X-ray emission from the same plasma has a different dependence on
the density, $S_x \propto \int n_e^2 \Lambda_{ee} dl$, where $\Lambda_{ee}$
is the X-ray cooling function. Making assumptions on the distribution
of the plasma (e.g., a $\beta$ profile) and taking advantage of the different
dependence on $n_e$, SZE and X-ray observations can be combined to determine
the distance to galaxy clusters.
These cluster distances  can be combined with
redshift measurements to determine the value of the Hubble constant
(Myers et al. 1997, Grainge et al. 2002, Reese et al. 2002).


In this paper we introduce a Markov chain Monte Carlo (MCMC)
procedure for the joint analysis of SZE and X-ray data.
The method is tested on the
galaxy cluster Abell 611 which has SZE data obtained with the Caltech
millimeter interferometric array at the Owens Valley Radio Observatory
(OVRO) outfitted with centimeter-wave receivers (Carlstrom, Joy and
Grego 1996) and X-ray data from the {\it Chandra} X-ray Observatory.
In a subsequent paper we will report the application of this technique
to a sample of $\sim$40 clusters.

\section{Data}

\subsection{{\it Chandra} X-ray data}
The Advanced CCD Imaging Spectrometer (ACIS) on board the {\it Chandra} X-ray Observatory provides 
high angular resolution  (half-power radius $\sim 0.5$ arcsec)
and good spectral resolution (50-300 eV FWHM) in the $\sim 0.3-10$ keV energy range.
We use the 
following steps to reduce the raw (Level 1) {\it Chandra} data:\\
(a) We use the `acis\_process\_events' tool from the Chandra Interactive
Analysis of Observations (CIAO) package to correct the Level 1 data
for charge transfer inefficiency.\\
(b) We generate a Level 2  event file applying standard
filtering techniques: we select grade=0,2,3,4,6, status=0 events and
filter the event file for periods of poor aspect solution using the
good time interval (GTI) data.\\
(c) Periods of high background count rates are occasionally present, typically due to
Solar flares (Markevitch 2001).
We discard these periods by constructing a light-curve
of a detector region devoid of astronomical sources, using a bin length of 500 seconds.
Time intervals that are in excess of the median count rate by more than 4$\sigma$ are  considered to
be affected by high background levels, and are discarded from the dataset.
This method is similar to that employed by Markevitch (2001) for the study of blank-sky
fields. \\
(d) We extract the cluster spectrum out to a radius
that encompasses 95\% of the cluster counts.
The 95\% radius is determined by extracting counts 
in annuli around the cluster
center, and forming the cumulative distribution.
All detectable point sources in the X-ray image are excluded.

Background in the ACIS instrument includes  detector and astronomical
components (see Markevitch et al. 2003). The background is particularly time-variable  in the
lowest energy channels (E$\leq$ 0.7 keV; Snowden et al. 1998). At the highest energies, the cluster emission
decreases and the signal becomes background-dominated. For these reasons, we
limit our analysis to the 0.7-7 keV range; this band includes the Fe complex lines at $\sim$ 6.7 keV
(in the rest frame) which are necessary for an accurate determination of the
plasma metallicity.

\subsubsection{Chandra observations of Abell~611 \label{chandra_obs}}
The X-ray data were obtained with the ACIS-S detector on the {\it Chandra} observatory
on Nov. 3, 2001 (OBS ID 3194), with a total on-source time of 36,599 s. 
The detector operating temperature was -120 C.
Periods of high background and poor aspect solution were filtered out, resulting in a
total effective exposure time of 36,114 s.

\begin{figure}
        \includegraphics[angle=-90,width=6in]{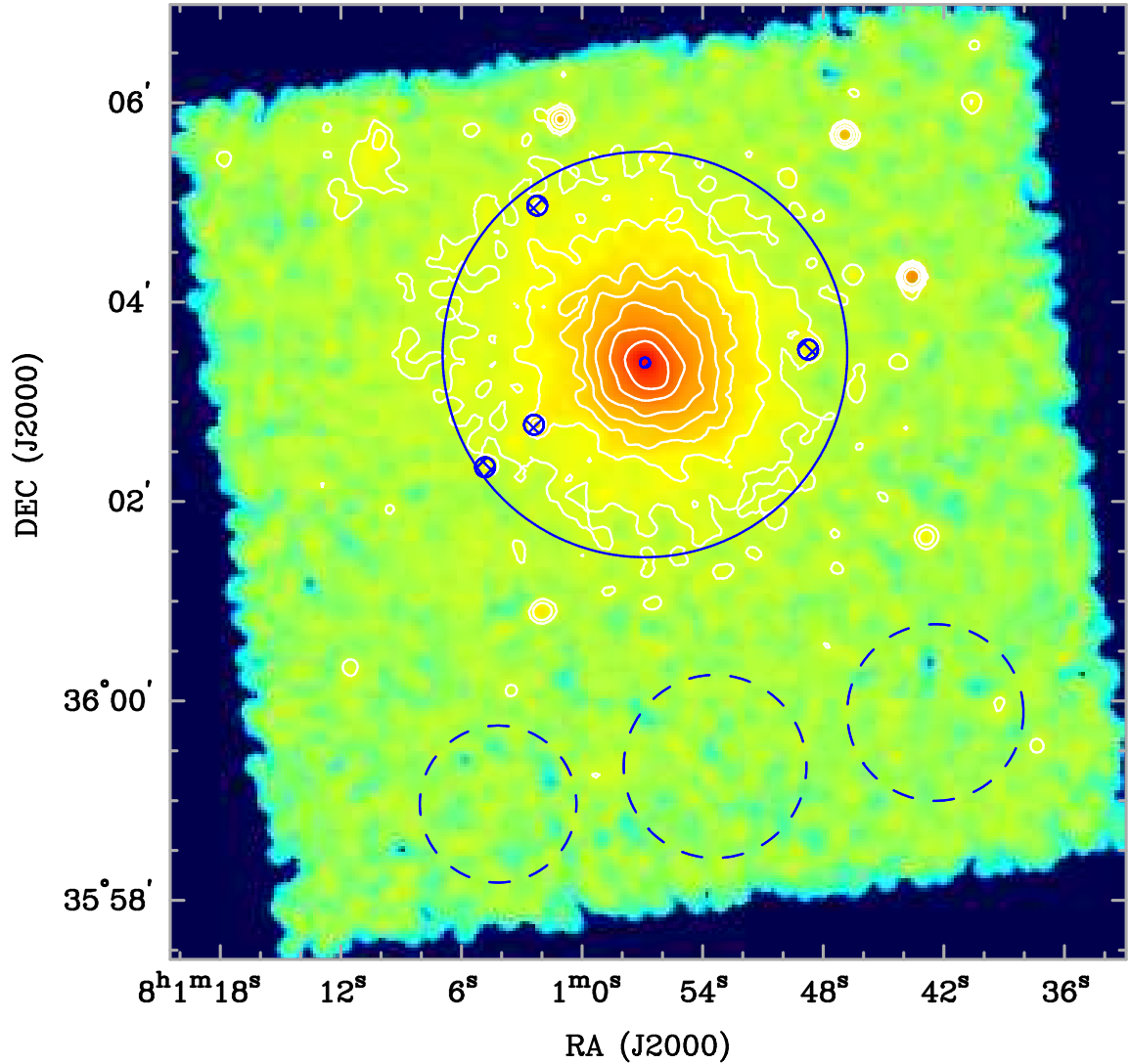}
\caption{{\it  Chandra} 0.7-7 keV image of Abell~611. Image is not exposure corrected,
contours are 0.5, 1, 2, 3, 5,  9 and 15 counts/pixel, pixel size is  1.97 arcsec.
The image was smoothed with a
Gaussian kernel of $\sigma$=4 arcsec.
The solid circle encompasses 95\% of the cluster counts and cross-hatched areas
were excluded due to point source contamination. Dashed circles
are the regions used for background determination.\label{a611_xray}}
\end{figure}

In Fig. \ref{a611_xray} we show the ACIS-S3 image of the cluster, smoothed with
a Gaussian kernel of $\sigma$=4 arcsec. The blue solid circle is the region used for
the spectral extraction, cross-hatched areas are the excluded regions and the blue dashed
circles are the regions used for background determination. Among the excluded regions is
the cD galaxy 2MASX~J08005684+3603234 (Crawford et al. 1999).
The X-ray background level was 8.6$\times 10^{-6}$ counts cm$^{-2}$ arcmin$^{-2}$ s$^{-1}$.
We assume a Galactic HI column density of $N_H=5.0\times 10^{20}$ cm$^{-2}$ (Dickey and
Lockman 1990) in the spectral analysis of the X-ray data.

\subsubsection{Background subtraction}

We investigate the ACIS background through the analysis of two
collections of blank-sky exposures
(acis57D2000-01-29bkgrndN0003.fits and acisiD2000-01-29bkgrndN0002.fits) provided
with the CIAO software for the purpose of background
estimation. The two datasets have exposure times of respectively 54 ks and 450 ks.
Several other blank-sky observations are available which cover the entire life span of the {\it Chandra}
mission.

\begin{figure}
      \includegraphics[angle=-90,width=12in]{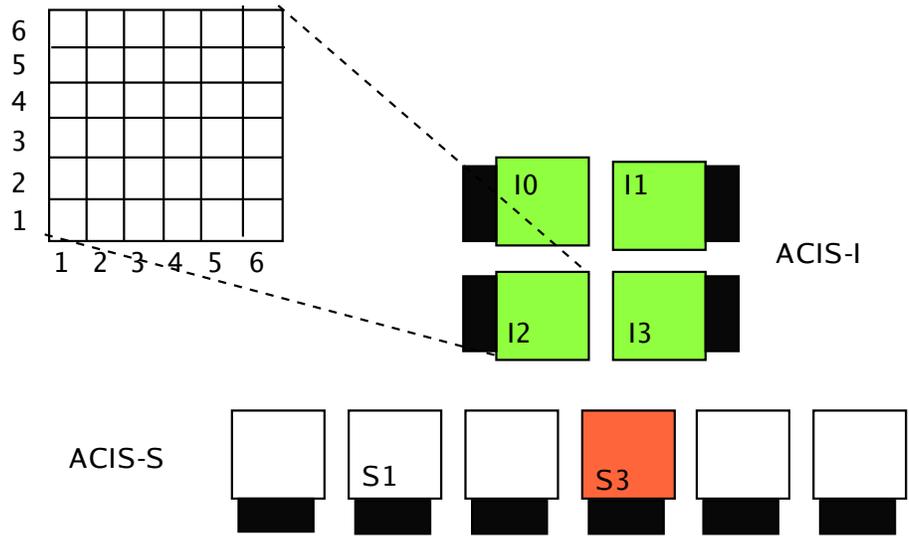}
      \vspace{-4in}	
      \caption{Layout of the ACIS detector, not to scale. Each CCD
is divided into 36 zones to study the spatial
behavior of the response. \label{layout}}
\end{figure}

The ACIS instrument is comprised of 10 CCDs on the focal plane
of the {\it Chandra} telescope. A layout of the ACIS instrument is
shown in Fig. \ref{layout} (see the {\it Chandra} Proposer's Observatory Guide
for further details). For each of the S3, I0, I1, I2, and I3 CCDs,
we select only events in the 0.7-7 keV energy range from the two
blank-sky observations. 
The linear size of each CCD is approximately 8 arcmin.
We divide each CCD into a 6x6 grid, as indicated in Fig.~\ref{layout}, to 
investigate the spatial variation of each CCD's response.

The S3 CCD has small spatial fluctuations in the detected counts, with a
standard deviation for the 36 regions that is 7.1~\% of the mean. 
We conclude that the ACIS-S3 CCD has a flat response, and that the
background can be simply estimated from a portion of the CCD which is devoid
of celestial sources (e.g., the galaxy cluster or other point sources).

\begin{figure}
      \includegraphics[angle=0,width=8in]{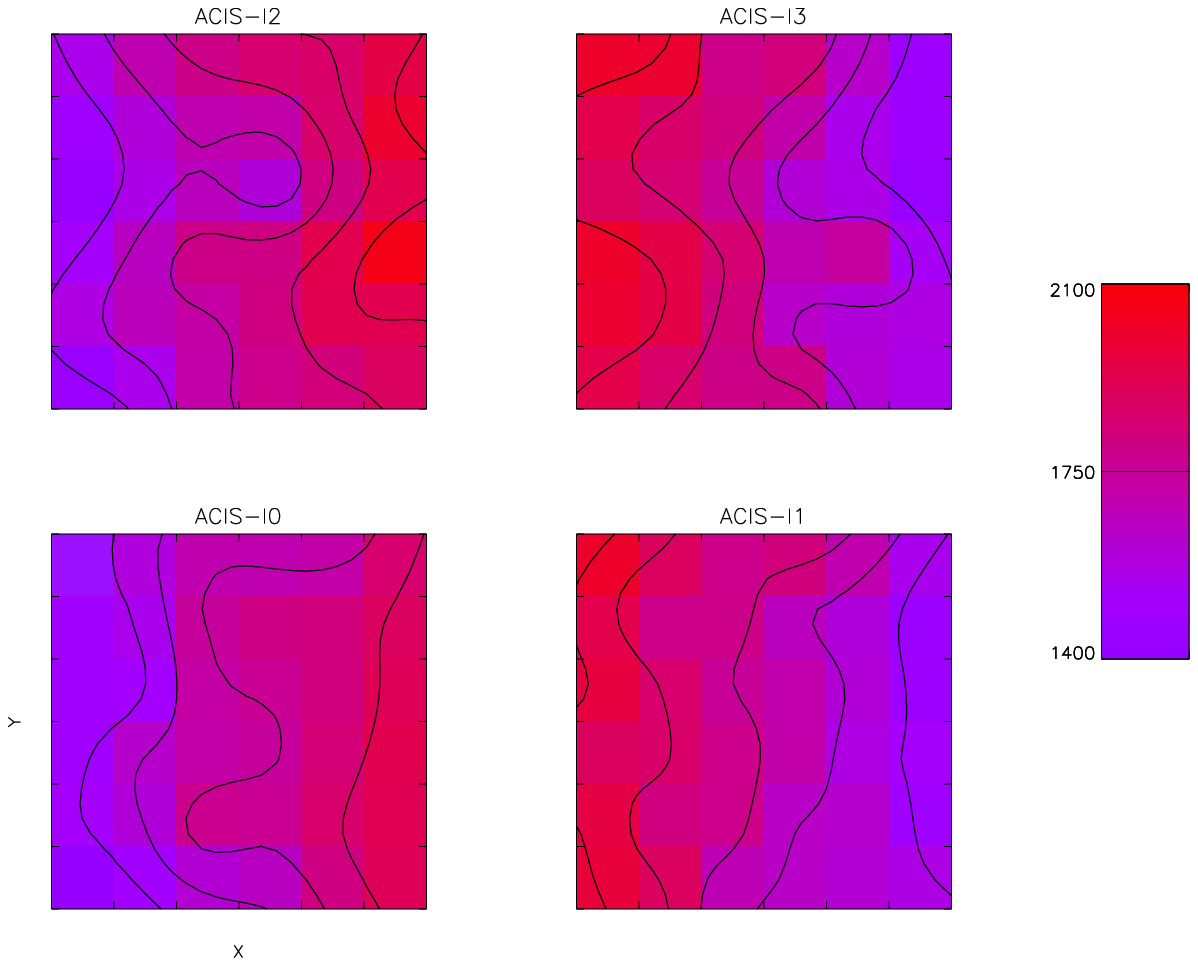}
      \caption{Spatial behavior of the ACIS-I response in 0.7-7 keV band. Counts
from the 450 ks blank-field exposure (see text for details) were accumulated in
$\sim 1.4 \times 1.4$ square arcmin pixels. \label{acisi}}
\end{figure}

\begin{figure}
        \includegraphics[angle=0,width=10in]{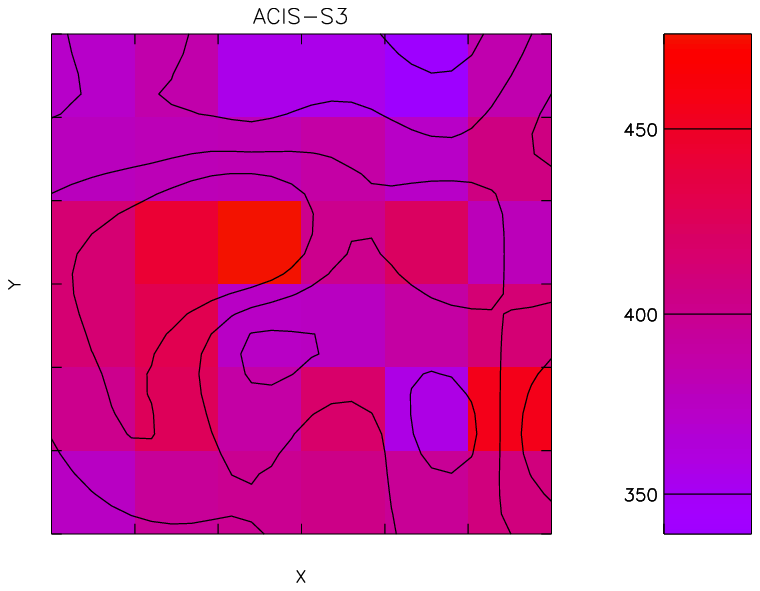}
      \caption{Spatial behavior of the ACIS-S3 response in 0.7-7 keV band. Counts
from the 54 ks blank-field exposure (see text for details) were accumulated in
$\sim 1.4 \times 1.4$ square arcmin pixels.
 \label{aciss3}}
\end{figure}

The situation is different for the ACIS-I CCDs. 
In all four ACIS-I CCDs the response increases with distance from the
read-out nodes, depicted as black rectangles in Fig. \ref{layout}.
This gradient in the response is at the level of 25~\%, and it results in a
standard deviation of about 10~\% of the mean of the counts for all CCDs, as
shown in Fig. \ref{acisi}. No such gradient is present in the S3 data, as shown in
Fig. \ref{aciss3}.
Similar results were found by Markevitch (2001) using other blank-field {\it Chandra} exposures.
Since the gradient and the absolute response are similar in all four ACIS-I CCDs,
the background can be estimated by using any (or all) of the
CCDs that are not contaminated by the cluster emission.
The aimpoint of the ACIS-I observations is on CCD I3, and CCDs I0, I1 and I2 are typically free of
cluster emission.

Our background estimation technique therefore
consists of selecting a region devoid of astronomical sources
from the same cluster observation, according to the following scheme:\\
(a) If the observation is performed with the ACIS-S configuration, the background is
chosen from peripheral regions of the same ACIS-S3 CCD  where the cluster is detected.\\
(b) If the observation uses the ACIS-I configuration, the background is chosen from the
3 CCDs (I0,I1 and I2) that are near the I3 CCD where the cluster aimpoint is located.
The background region is at the same distance from the read-out nodes
as the cluster. This is the case of several observations
to be presented in  follow-up papers.


\subsection{Interferometric SZE data}
\begin{figure}
        \includegraphics[angle=-90,width=6in]{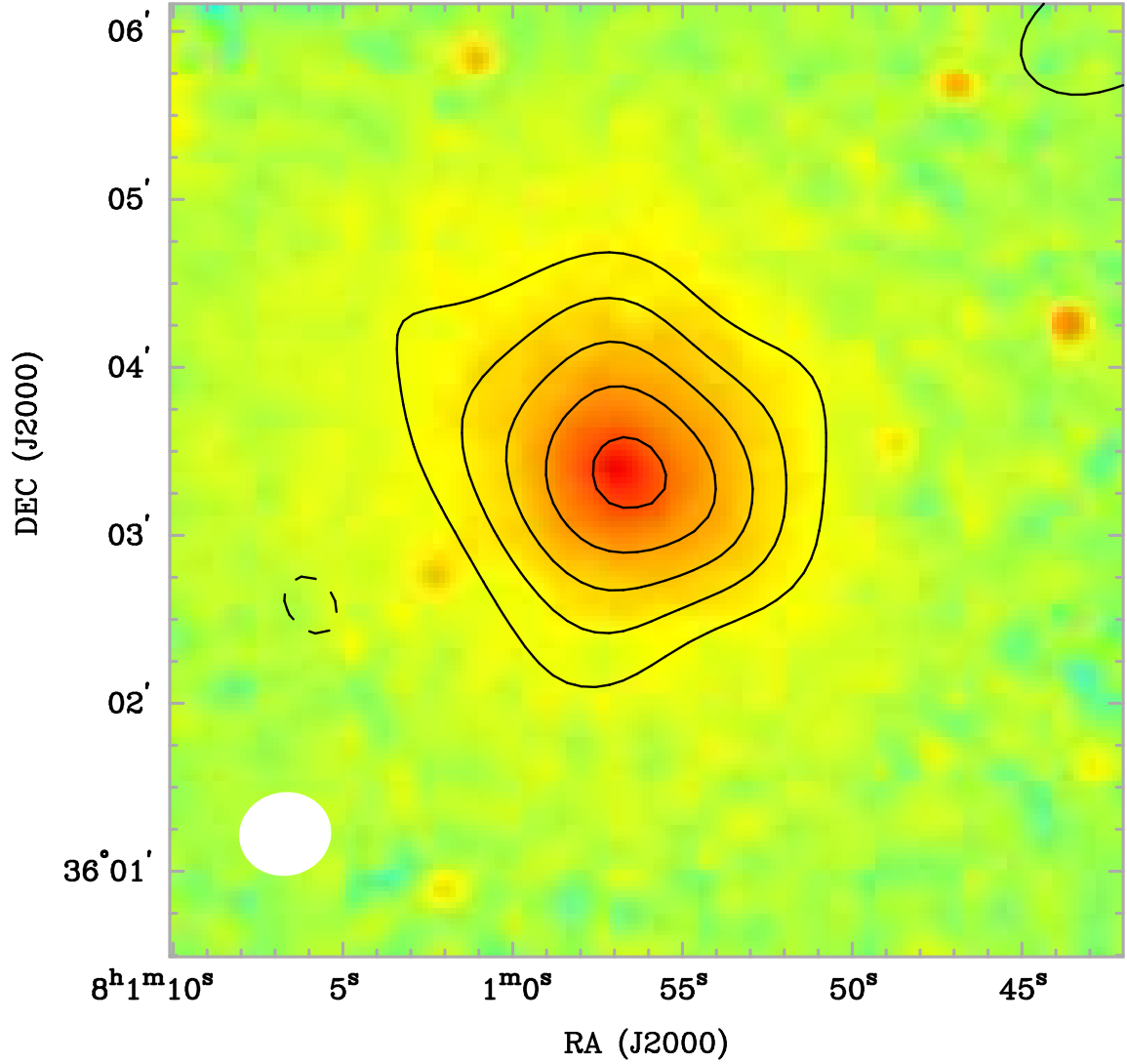}
\caption{SZE contours overlaid on the {\it Chandra} 0.7-7 keV image of Abell~611. 
The rms noise is 40 $\mu$Jy/beam and the
contours are at -5, -4, -3, -2, -1 and 1$\times 112$ $\mu$Jy/beam; negative (positive)
contours are shown as solid (dashed) lines.
The synthesized beam is 62x56 arcsec.
\label{a611_SZE}}
\end{figure}

The SZE measurements discussed in this paper were obtained by
outfitting the 6-element OVRO millimeter array with centimeter
wavelength receivers (Carlstrom, Joy and Grego 1996). The details of
the observations and data reduction are covered extensively in Reese
et al.\ (2002) and only reviewed briefly here.  The data were taken
with two 1 GHz wide bands centered at 29 GHz and 30GHz with receiver
temperatures $T_{rx} \sim$ 11-20 K.  The typical system temperatures
scaled to above the atmosphere were $T_{sys} \sim$ 45 K.  Multiple
configurations of the telescopes were used which typically placed most
of the telescopes in a compact configuration to maximize the
sensitivity on the angular scale of distant galaxy clusters ($\sim$ 1
arcmin), but with a subset of the telescopes forming longer baselines
to provide simultaneous detection (and subsequent removal) of
radio background point sources.  
Abell~611 was observed with
the OVRO array for a total of 57 hours.
The SZE data were reduced using the MMA
software package (Scoville et al. 1993). 
No point sources were found
with a 3$\sigma$ upper limit of $135 \mu$Jy (not corrected
for attenuation by the $4.2'$ FHWM primary beam).  In
Fig. \ref{a611_SZE} we show a contour plot of the SZE data overlaid on
the Chandra X-ray image.

\section{Obtaining cluster distances \label{distance}}

The high computational efficiency of the MCMC analysis and the
improvements in the X-ray and SZE data will allow more complex models
for the instracluster gas.  To demonstrate the MCMC analysis, however,
we follow the procedure detailed
by Reese et al.\ (2002).
We assume that the spatial
distribution of the cluster plasma is described by a spherical
isothermal $\beta$-model (Cavaliere and Fusco-Femiano 1976, 1978) given
by:
\begin{equation}
n_e(r)=n_{e0} \left( 1+\frac{r^2}{r_c^2}\right)^{-\frac{3 \beta}{2}}
\end{equation}
where $n_e$ is the electron number density, $r$ is the radius from the cluster's center,
$r_c$ is the core radius and $\beta$ is a power-law index.
With this model, the thermodynamic SZE temperature decrement/increment $\Delta T$
is

\begin{equation}
\Delta T= f_{(x,T_e)} T_{CMB} D_A \int d \zeta \sigma_T n_e \frac{k_B T_e}{m_e c^2}=
\Delta T_0 \left( 1+\frac{\theta^2}{\theta_c^2} \right)^{\frac{1-3\beta}{2}}
\end{equation}
where $f_{(x,T_e)}$ is the frequency dependence of the SZE with $x=\frac{h \nu}{k_B T_{CMB}}$
(e.g., Reese et al. 2002 and references therein),
$D_A$ is the cluster angular diameter distance, $T_{CMB}$=2.728 K (Fixsen et al. 1996),
$\theta$ is the angular radius in the plane of the sky, $\theta_c$ the corresponding
angular core radius, $\sigma_T$ is the Thomson cross section, $k_B$ is Boltzmann's constant,
$c$ is the speed of light in vacuo, $m_e$ is the electron's mass and $T_e$ the electron's temperature.
The integration is performed along the line of sight, $l$, and we define $\zeta=l/D_A$.

The X-ray surface brightness is given by
\begin{equation}
S_X=\frac{1}{4 \pi (1+z)^4} D_A \int d \zeta n_e^2 \Lambda_{ee} = 
S_{X0} \left( 1 +\frac{\theta^2}{\theta_c^2} \right)^{\frac{1-6\beta}{2}} 
\end{equation}
where $S_X$ is the surface brightness,
$z$ is the cluster's redshift
and $\Lambda_{ee}=\Lambda_{ee}(T_e,A)$ is the X-ray cooling function of the gas.

Eq. 2 and 3 can be combined to determine $D_A$:
\begin{equation}
D_A=\frac{\Delta T_0^2}{S_{X0}} \left( \frac{m_e c^2}{k_B T_{e0}} \right)^2 
\frac {\Lambda_{ee}}{4 \pi f^2_{(x,T_e)} T^2_{CMB} \sigma^2_T (1+z)^4}
\frac{1}{\theta_c} \left[ \frac{\Gamma(3 \beta/2)}{\Gamma(3 \beta/2 - 1/2)} \right]^2 \frac{\Gamma(3 \beta -1/2)}{\Gamma(3 \beta)}
\label{DA}
\end{equation}
where $\Gamma()$ is the gamma function which comes from
the integration of the $\beta$-model along the central line of sight. 
Similarly, one can eliminate $D_A$ to
obtain $n_{e0}$ (e.g., Reese et al. 2002; Birkinshaw  1999).

\section{Joint SZE and X-ray data modelling \label{model2}}
The SZE and the X-ray emission both depend on the properties of the hot
cluster plasma. 
We use a joint model for the interferometric SZE radio data and the {\it Chandra} X-ray data 
which describes all the relevant spatial and spectral characteristics
necessary for the distance measurement.
The model consists of:\\
(a) A $\beta$-model of the plasma distribution (see Eq. 2 and 3). The model includes the variable parameters
$S_{X0}$, $\beta$, $r_c$, $\Delta T_0$,  and the fixed coordinates of the cluster center.\\
(b) A photoabsorbed optically-thin spectral model (Raymond-Smith and WABS
models in XSPEC) to determine $T_e$ and metal abundance $A$ of the gas.
Solar metal abundances
are from Feldman (1992). The cluster plasma is assumed
to be isothermal.\\
(c) Additional parameters: the X-ray background level, held fixed at its measured 
value (see sec. \ref{chandra_obs}) and, if present, the position and flux
of radio point sources. No radio point sources are found in the SZE data for A611.
 
In summary, the model  is described by a set of
parameters \bmath{\theta} $\equiv$($\beta$, $r_c$, $\Delta T_0$, $S_{X0}$, $T_e$, $A$).
For each parameter set \bmath{\theta}, we calculate the joint likelihood
$\mathcal{L}$ (e.g., Bevington 1969) of the model with the available data. Since
the datasets (SZE and X-ray) are independent, the joint likelihood
is the product of the individual likelihoods. 

The interferometric
SZE data provide constraints in the Fourier u-v plane, and the likelihood of the
SZE dataset is therefore directly calculated in u-v coordinates
(see Reese et al. 2000). The likelihood of the X-ray images is
calculated pixel by pixel. For the spectral X-ray data,
an analytical model of the emissivity (available through the XSPEC software)
is convolved with the {\it Chandra} response, and the likelihood is calculated in each
spectral channel.

We employ the computationally
efficient Markov chain Monte Carlo  method in order to handle the large numbers of
parameters involved in the joint SZE/X-ray spatial and spectral
analysis.

\subsection{The Markov chain method}
The Markov chain Monte Carlo method can be used to obtain the
probability distribution function of model parameters based on observational data
(Gamermann 1997, Gilks, Richardson and Spiegelhalter 1996, Christensen and Meyer 2001, 
Lewis and Bridle 2002, MacKay 2003,
Marshall, Hobson and Slosar 2003, Hobson and McLachlan 2003, Verde et al. 2003, Christensen et al. 2001).
To create the Markov chain, we choose candidate parameter values ($\bmath{\theta^{\prime}}$)
from the allowed parameter space (the parameter support, Table \ref{model}).
The support values in Table  \ref{model} were determined from test 
runs of the Markov chain with very broad parameter limits
to ensure that all 
statistically acceptable regions of parameter space were included.

Given these candidate parameters, we compute the likelihood of the model and
the candidates are either accepted or rejected based on the Metropolis-Hastings 
acceptance algorithm (Metropolis
et al. 1953; Hastings 1970; Gilks, Richardson and Spiegelhalter 1996).
When a sufficiently large number of parameters has been accepted into the Markov
chain, the frequency of their occurrence in the chain approaches the true 
probability distribution function (Gamermann 1997, Gilks, Richardson and Spiegelhalter 1996,
MacKay 2003).

 

\subsection{Proposal distribution and starting point}

In constructing a MCMC one has the freedom to choose
any desired method for drawing candidates (Roberts 1996; MacKay 2003).
We draw candidates $\bmath{\theta^{\prime}}$  that are in the neighborhood
of the previously accepted parameters, instead of drawing them from the
full parameter space. Our `proposal' distribution is
a simple top-hat function.
A small width for the proposal distribution typically results in
a high acceptance rate, as candidates have a likelihood that is
similar to that of the previously accepted parameters.
On the other hand,
a large number of steps is required for the Markov chain to sample the entire parameter space.
We tested proposal distribution widths from 10\% to 50\% of the parameter support (see Table \ref{model}),
calculated the number of iterations required for convergence (see the following section)
and present the results in Table \ref{numiter}.
The optimum proposal distribution width is approximately 25\%, and we use this value
for all subsequent analysis. 
                                                                                                                                                            
The starting point of a Markov chain can be chosen arbitrarily (e.g., Gilks, Richardson and Spiegelhalter 1996).
We start the Markov chain at the midpoint of each parameter's support (see Table \ref{model}), and
run the Markov chain for 100,000 iterations. The acceptance rate for a 25\% proposal distribution width is 11.4\%.

\subsection{Convergence of the Markov chain}

A Markov chain requires a large number of steps before
it reaches convergence. We test the convergence
using three independent tests: the Raftery-Lewis diagnostic (Raftery and Lewis 1992),
the Gelman-Rubin diagnostic (Gelman and Rubin 1992) and the 
Geweke test (Geweke 1992; Gamermann 1997; Geyer 1992).

The Raftery-Lewis method determines whether convergence to the stationary distribution has been
achieved, and also estimates the number of iterations that are required 
in order to determine confidence intervals to a specified accuracy.
We used the CODA routines (Best, Cowles, and Vines, S.K. 1995,
Plummer et al. 2004) to compute the
Raftery-Lewis statistics, and present the results in Table \ref{converg}.
For all of the parameters, the number of iterations required is less
than our chain length of 100,000.

\begin{figure}
        \includegraphics[angle=0,width=6in]{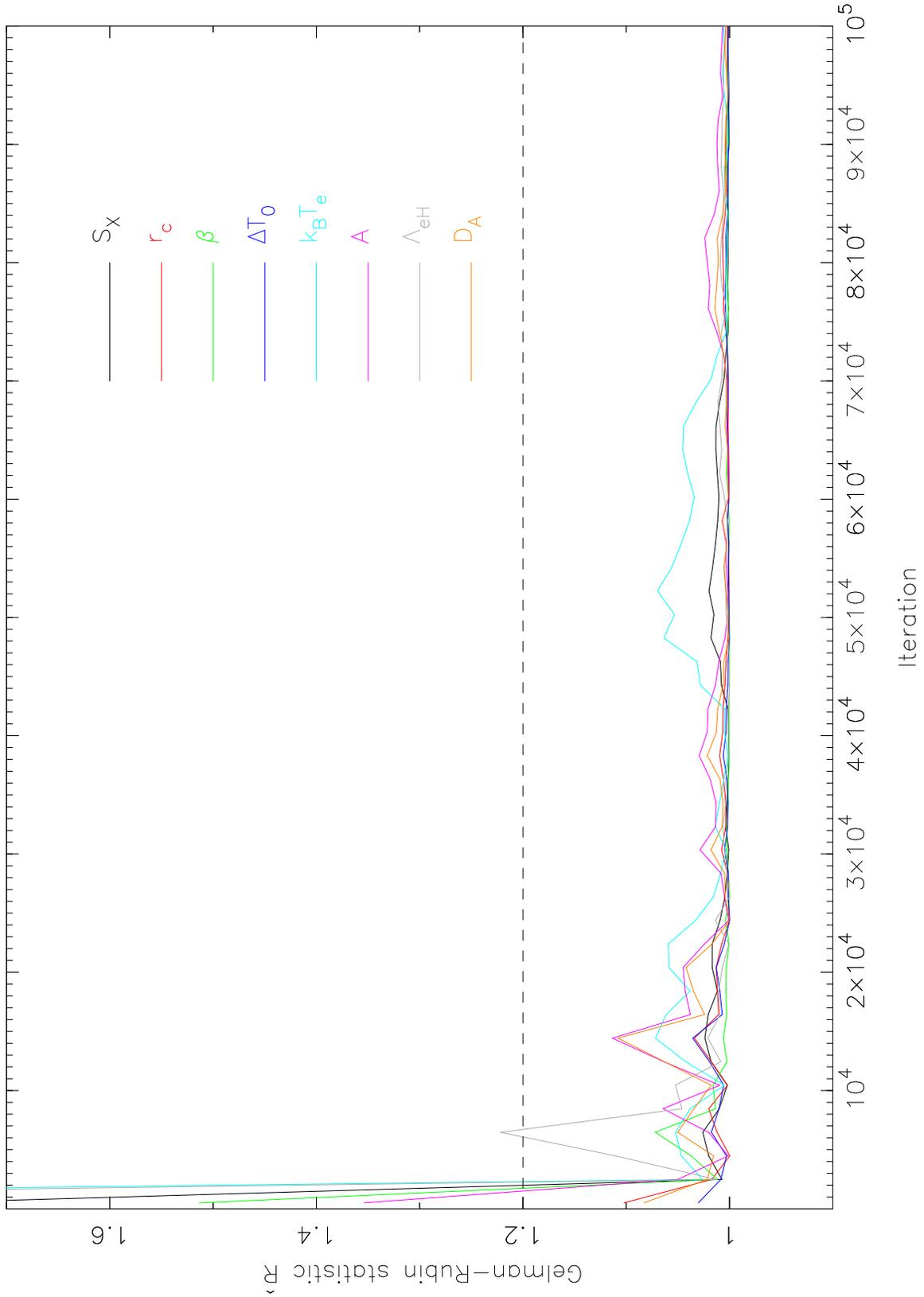}
\caption{Gelman-Rubin statistic $\hat{R}$ obtained from 3 simultaneous Markov chains
with different initial conditions. \label{gr}}
\end{figure}

The Gelman-Rubin test is based on several parallel Markov chains, each started from different initial values. 
The method calculates a factor, $\hat{R}$, based on the variance within and between each chain.
Convergence is indicated by $\hat{R} \lesssim$ 1.2 (Gelman 1996).
Fig. \ref{gr} shows the Gelman-Rubin statistic computed from three parallel chains; the test
indicates that convergence is achieved early in our 100,000 element chain.

Finally, we compute the Geweke $z_G$ statistic.
We discard the initial $n_i$=5,000 steps of the chain (the {\it burn-in} period),
 and divide the remaining
$n$=95,000 steps into the initial 10\% ($n_b=9,500$) and the final 50\%
($n_a=47,500$) segments,
according to Geweke (1992). The intermediate 40\% portion in not used
for testing the convergence.
Since the values in the Markov chain are correlated by construction, the initial
and final segments are averaged over 100 steps, to ensure that the rebinned
values are uncorrelated (Roberts 1996). 
The $z_G$ function, also known as the Geweke z-score, is the standardized difference between the initial 
and the final portions of the chain, and it
is distributed as a standard Gaussian N(0,1) if the chain
has reached convergence.
Convergence is therefore indicated if $z_G$ values are less than  $\sim 3 \sigma$ for all
parameters (Geweke 1992; Gamerman 1997). 
The $z_G$ values for our
Markov chain are shown in Table \ref{converg}, and are all within $\pm 2 \sigma$, consistent
with convergence to the stationary distribution.

\subsection{Results of the Markov chain Monte Carlo for Abell~611}

\begin{figure}
        \includegraphics[angle=0,width=6in]{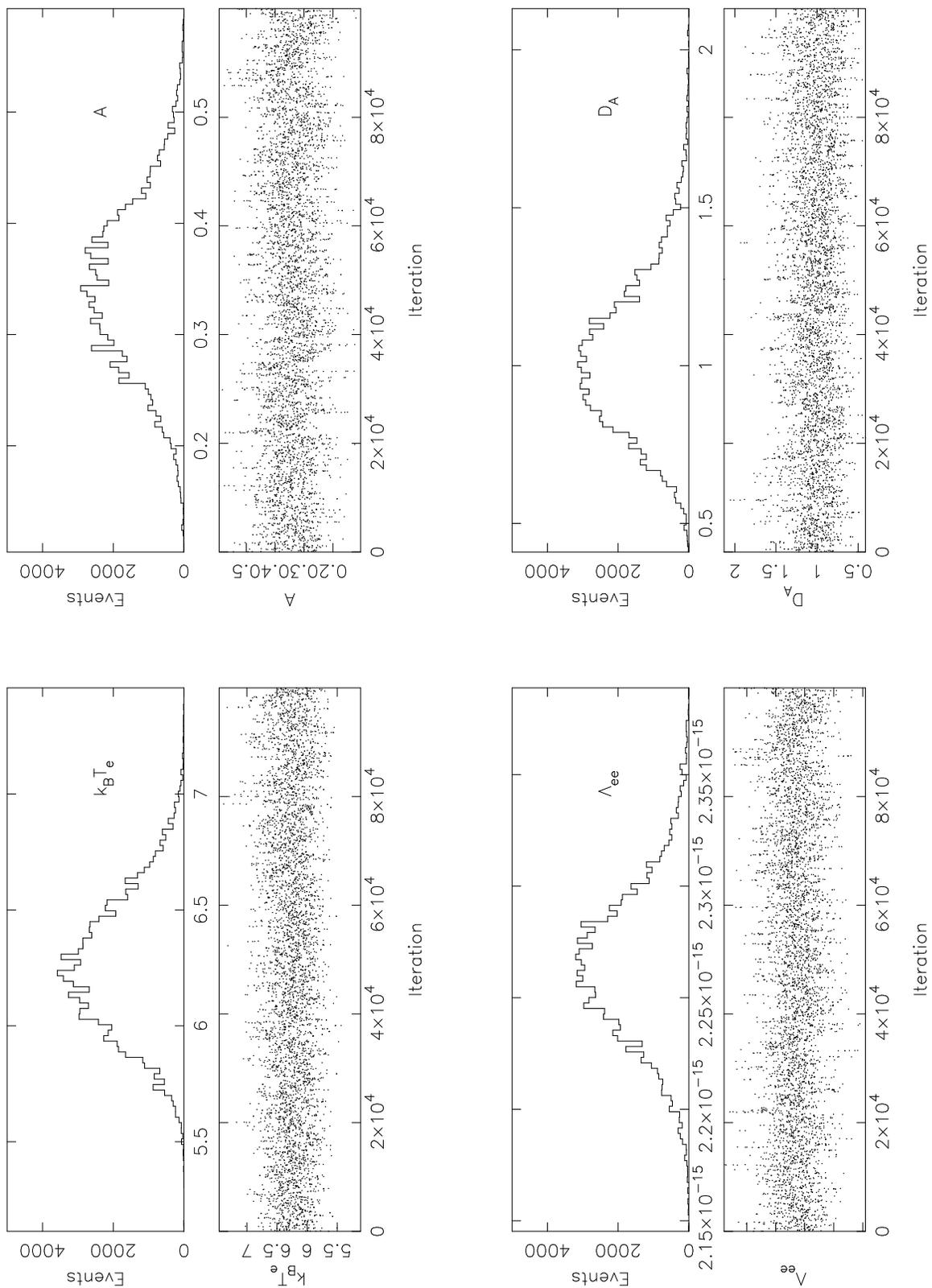}
\caption{Results of the Markov chain for all parameters
Top: Probability distribution. We excised the initial 5,000 events,
and used 95,000 events to determine the distribution.
Bottom: Time series of the Markov chain including all 100,000 events.\label{fig_results}}
\end{figure}

In Fig. \ref{fig_results} we  show the probability distribution functions for all parameters.
The 68\% and 90\% confidence intervals (calculated by marginalizing
over all other parameters) are given in Table \ref{results}.

Thinning the Markov chain is often employed to weed out values which are highly correlated with
neighboring elements in the chain. To examine the effect this would have on the derived confidence
intervals, we selected every tenth element in the chain and recalculated the statistics.
The results (right side of Table \ref{results}) indicate that there is no significant change in the derived
confidence intervals. Finally, we calculated the correlation coefficients between parameters, and the results
are shown in Table \ref{xcorr}. As expected, parameters $r_c$ and $\beta$ are strongly correlated
(see Reese et al. 2002, Grego et al. 2001).

\subsection{Comparison with other analysis methods}
We employ the XSPEC spectral code to compare the confidence intervals on the
spectral parameters $T_e$ and $A$. Table \ref{comparison} shows the comparison
between the MCMC-derived best-fit values and confidence intervals with those
provided by XSPEC.
The results are in excellent agreement, confirming our
MCMC results. 
In addition, we use the CIAO Sherpa software to determine the $\beta$-model parameters,
and results are also shown in Table \ref{comparison}.
The agreement with the results of Table \ref{results} provides additional
confidence in the reliability of our analysis.

Abell~611  data were also analyzed by Reese et al. (2002).
Using the same SZE data, but lower resolution 
ROSAT and ASCA X-ray data, they found a
SZE decrement of $\Delta T_0 = -853 \pm^{120}_{140}\; \mu$K. Our results
of $\Delta T_0 = -801 \pm{84}\; \mu$K are again in very good agreement with theirs.

\section{Conclusions}
We present a Markov chain Monte Carlo technique to derive cluster distances
from SZE and X-ray data.
The method was succesfully tested on the OVRO and {\it Chandra} data of Abell~611, a galaxy
cluster at z=0.288. We measure an angular diameter distance of $D_A=1.00\pm^{0.24}_{0.21}$ Gpc
(68\% confidence level). In a previous work based on the same SZE data, but using 
lower-resolution X-ray data, Reese et al. (2002) derived a distance of $D_A=0.99\pm^{0.32}_{0.29}$
Mpc. 

The {\it Chandra} X-ray data provides
simultaneous spatial and spectral information, featuring the finest angular resolution
to date ($\sim 0.5$ arcsec half-power radius).
The MCMC method has two major advantages: 
it is computationally
more efficient than the traditional likelihood ratio-based methods
and it provides simultaneously the probability distribution function
of all model parameters.

This technique will be used in future papers to determine the distances of a large sample of
galaxy clusters for which there are available high-resolution {\it Chandra} X-ray data
and BIMA/OVRO SZE data.

This work is supported by NASA LTSA grant NAG 5-7986. E.D.R. acknowledges support from
NASA Chandra Postdoctoral Fellowship PF 1-20020. 
Partial support was also provided by NSF grants
AST-0096913 and PHY-0114422. S.J.L.\ acknowledges support from NASA GSRP Fellowship  NGT5-50173.
We thank the referee for  helpful
comments and suggestions.

\newpage

\begin{deluxetable}{lcccc}
\tabletypesize{\small}
\tablecaption{Parameters in the Markov chain Monte Carlo \label{model}}
\tablehead{Parameter & Starting value & Lower limit & Upper limit &  Units }
\startdata
$S_{X0}$ &  1.02 & 0.89 & 1.16 &  $10^{-3}$ counts s$^{-1}$ cm$^{-2}$ arcmin$^{-2}$\\
$r_c$ & 20 &  17 & 23 &  arcsec\\
$\beta$ & 0.59 &  0.56 & 0.62 & --\\
$\Delta T_0$ & -0.80 &  -1.10 & -0.50 &   mK\\
$k_B T_e$ & 6.25 &  5 & 7.5 &   keV \\
$A$     &  0.35  & 0.1 & 0.6 &  solar \\
\enddata
\end{deluxetable}

\begin{deluxetable}{lrrrrr}
\tabletypesize{\small}
\tablecaption{Number of iterations required for convergence as a function of proposal distribution width\label{numiter}}
\tablehead{ Parameter & \multicolumn{5}{c}{Proposal distribution width} \\
                &     \multicolumn{5}{c}{\hrulefill} \\
                      & 10\%  & 20\%  & 25\% & 30\% & 50\% }
\startdata
$S_{X0}$       &113680 & 60885 & 29190 & 50256 & 91122  \\
$r_c$       &165540 & 68498 & 86906 & 47957 & 101375\\
$\beta$     &118503 & 64832 & 73780 & 61800 & 97292\\
$\Delta T_0$&101200 & 73270 & 40185 & 50892 & 97379\\
$k_B T_e$   &68180  & 74148 & 32916 & 92579 & 99748\\
$A$         &129903 & 49358 & 89360 & 38143 & 89423\\
$\Lambda_{ee}$ &99160 & 59080 & 65918 & 52810 & 100557\\
$D_A$       &88480  & 67312 & 62899 & 53478 & 106959\\
\enddata
\tablecomments{The number of iterations required for convergence are obtained from the Raftery-Lewis
test, for a 68\% confidence interval (see Table \ref{converg}).}
\end{deluxetable}

\begin{deluxetable}{lccccc}
\tabletypesize{\small}
\tablecaption{Markov chain Convergence Tests \label{converg}}
\tablehead{  & \multicolumn{2}{c}{Raftery-Lewis} & \multicolumn{2}{c}{Raftery-Lewis} & Geweke \\
         &   \multicolumn{2}{c}{(68\% C.I.)} & \multicolumn{2}{c}{(90\% C.I.)} & \\
         &  \multicolumn{2}{c}{\hrulefill} & \multicolumn{2}{c}{\hrulefill} & \hrulefill\\
        & Burn-in  & Iterations required &  Burn-in  & Iterations required & $z_G$ \\
Parameter        & Iterations & for convergence & Iterations & for convergence & }
\startdata
$S_{X0}$       & 108  &   29190  &  97   &   36518 & 0.78\\
$r_c$       & 304  &   86906  & 225   &   78690 & -0.84\\
$\beta$     & 280  &   73780  & 138   &   51794 & -0.80\\
$\Delta T_0$& 147  &   40185  & 137   &   51595 & 1.45\\
$k_B T_e$   & 122  &   32916  &  95   &   35775 & 0.34\\
$A$         & 320  &   89360  & 171   &   67583 & -1.12\\
$\Lambda_{ee}$ & 253&  65918  & 160   &   58176 & -1.03\\
$D_A$       & 248  &   62899  & 109   &   40924 & -1.59\\
\enddata
\tablecomments{We used the CODA software to calculate the Raftery-Lewis statistics.
For the 68\% confidence interval (corresponding to
the q=0.16 quantile), we required an accuracy of r=2\% with a probability of s=90\%. For the 90\% confidence interval
(q=0.05 quantile),
we required an accuracy of r=1\% with a probability of s=90\%. }
\end{deluxetable}

\newpage                                                                                                                                                            
\begin{deluxetable}{lcccccc}
\tabletypesize{\small}
\tablecaption{Markov chain Monte Carlo results for Abell~611 data \label{results}}
\tablehead{
 & \multicolumn{3}{c}{FULL CHAIN} & \multicolumn{3}{c}{THINNED CHAIN}\\
 & \multicolumn{3}{c}{\hrulefill} & \multicolumn{3}{c}{\hrulefill} \\
Parameter & Median & 68 \% interval & 90 \% interval & Median & 68 \% interval & 90 \% interval }
\startdata
$S_{X0}$ & 1.02 & $\pm^{0.03}_{0.03}$ & $\pm^{0.05}_{0.05}$ & 1.02 & $\pm^{0.03}_{0.03}$& $\pm^{0.05}_{0.05}$ \\
$r_c$ & 20.00 & $\pm^{0.70}_{0.62}$  & $\pm^{1.14}_{1.10}$  & 20.00 & $\pm^{0.69}_{0.61}$  & $\pm^{1.14}_{1.02}$   \\
$\beta$ & 0.594 & $\pm^{0.008}_{0.007}$ & $\pm^{0.013}_{0.012}$ & 0.594 &  $\pm^{0.006}_{0.007}$ & $\pm^{0.013}_{0.012}$   \\
$\Delta T_0$ & -0.801 & $\pm^{0.084}_{0.083}$ & $\pm^{0.133}_{0.139}$ & -0.800 & $\pm^{0.082}_{0.082}$ & $\pm^{0.132}_{0.140}$ \\
$k_B T_e$ & 6.25 & $\pm^{0.30}_{0.28}$ & $\pm^{0.50}_{0.43}$ & 6.25 & $\pm^{0.30}_{0.28}$ & $\pm^{0.50}_{0.42}$  \\
$A$ & 0.34 & $\pm^{0.07}_{0.07}$ & $\pm^{0.12}_{0.12}$ & 0.34 & $\pm^{0.07}_{0.07}$ & $\pm^{0.07}_{0.07}$  \\
$\Lambda_{ee}$ & 2.26 & $\pm^{0.03}_{0.03}$ & $\pm^{0.06}_{0.05}$ & 2.26 & $\pm^{0.03}_{0.03}$ & $\pm^{0.06}_{0.04}$   \\
$D_A$ & 1.00 & $\pm^{0.24}_{0.21}$ & $\pm^{0.43}_{0.33}$ & 1.01 & $\pm^{0.24}_{0.21}$ & $\pm^{0.42}_{0.33}$ \\
\hline
\tablecomments{Units of measure for $\Lambda_{ee}$ are counts cm$^{3}$ s$^{-1}$, and $D_A$ is measured in Gpc.
For units of measure of all other quantities, see Table \ref{model}.}
\enddata
\end{deluxetable}

\newpage
\begin{table}
\caption{Correlation coefficients for parameters in Markov chain Monte Carlo \label{xcorr}}
$\begin{array}{r|rrrrrrrr}
 &  S_{X0}  & r_c & \beta & \Delta T_0 & A &k_B T_e &  \Lambda_{ee}  & D_A\\
 \hline 
S_{X0} &       1.00 &-0.86 &-0.64 &-0.10 & 0.02 & 0.03 & 0.03 &-0.02\\
r_c &               & 1.00 & 0.93 & 0.14 &-0.01 &-0.04 &-0.02 &-0.01\\
\beta &             &      & 1.00 & 0.15 &-0.00 &-0.04 &-0.02 &-0.03\\
\Delta T_0&         &      &      & 1.00 &-0.01 &-0.00 &-0.01 &-0.90\\
A &                 &      &      &      & 1.00 &-0.15 & 0.93 & 0.13\\
k_B T_e &           &      &      &      &      & 1.00 & 0.22 &-0.40\\
\Lambda_{ee}&       &      &      &      &      &      & 1.00 &-0.02\\
D_A &               &      &      &      &      &      &      & 1.00\\
\end{array}$
\end{table}

\begin{deluxetable}{lcc}
\tabletypesize{\small}
\tablecaption{Comparison of Markov chain Monte Carlo results with XSPEC and CIAO Sherpa results \label{comparison}}
\tablehead{Parameter & 68\% interval & 68\% interval\\
                     & (MCMC)        & (XSPEC/Sherpa) }
\startdata
$k_B T_e$ & 6.23$\pm^{0.29}_{0.29}$  & 6.20$\pm{0.30}$ (XSPEC) \\
$A$ & 0.34$\pm^{0.07}_{0.07}$  & 0.35$\pm^{0.07}_{0.08}$ (XSPEC)\\
$S_{X0}$ & 1.02$\pm^{0.03}_{0.03}$& 1.07$\pm^{0.03}_{0.03}$ (Sherpa)\\
$r_c$ &20.00$\pm^{0.70}_{0.62}$ & 19.42$\pm{0.63}$ (Sherpa)\\
$\beta$ & 0.594$\pm^{0.008}_{0.007}$  & 0.589$\pm{0.01}$ (Sherpa)\\
\enddata
\tablecomments{The CIAO Sherpa spatial fit is based only on the {\it Chandra} X-ray data, not the
combination of X-ray and radio data, as in Table \ref{results}.
We determined, however, that the results of Table \ref{results} are
virtually unchanged when the X-ray data alone are used for the
determination of the beta model; the high S/N X-ray data drive the fit of
the spatial parameters.
}
\end{deluxetable}
\newpage                                                                                                                                                            


\begin{figure}
	\includegraphics[angle=0,width=6in]{f7a.eps}       
\end{figure}

\end{document}